\documentstyle[9pt,emulateapj]{article}

\received{}
\revised{}
\accepted{}
\journalid{}{}
\articleid{}{}
\paperid{}
\cpright{}{}
\ccc{}

\slugcomment{Submitted to the Astrophysical Journal}

\lefthead{Wood et al.}
\righthead{Optical Variability of HH30}

\begin{document}

\title{Optical Variability of the T~Tauri Star HH30\altaffilmark{1}}

\author{Kenneth Wood\altaffilmark{2}, Scott J. Wolk\altaffilmark{2}, 
K.Z. Stanek\altaffilmark{2,3}, George Leussis\altaffilmark{2}, 
Keivan Stassun\altaffilmark{4}, \\
Michael Wolff\altaffilmark{5}, 
Barbara Whitney\altaffilmark{5}}

\altaffiltext{1}{Based on observations obtained with the
1.2 m Telescope at the F. L. Whipple Observatory}

\altaffiltext{2}{Harvard-Smithsonian Center for Astrophysics, 
60 Garden Street, Cambridge, MA~02138;\\
kenny@claymore.harvard.edu, swolk@cfa.harvard.edu, gleussis@cfa.harvard.edu, 
kstanek@cfa.harvard.edu}

\altaffiltext{3}{Hubble Fellow}

\altaffiltext{4}{Astronomy Department, University of Wisconsin, 475 North 
Charter Street, Madison, WI~53706; keivan@astro.wisc.edu}

\altaffiltext{5}{Space Science Institute, 3100 Marine Street, Suite A353, 
Boulder, CO~80303; bwhitney@colorado.edu, wolff@colorado.edu}

\authoremail{kenny@claymore.harvard.edu}

\begin{abstract}

We report results of $VRI$ photometric monitoring  of the T~Tauri 
star plus disk system HH30.  
We find that HH30 is highly variable over timescales of a few days 
with $\Delta V\sim 1.5$ mag, $\Delta I\sim 1.1$ mag.  
Furthermore we find hints of periodicity with periodograms indicating 
possible periods of 11.6 and 19.8~days.  
The $VRI$ photometry is available through the {\tt anonymous ftp} service.  
We model 
the variability with Monte Carlo radiation transfer simulations for 
a spotted star plus disk system and find that the large variability 
is best reproduced with a single hot spot 
and circumstellar grains that are larger than typical interstellar grains.  
The apparent existence of a single hot spot 
and the need for larger, more forward throwing grains is consistent 
with previous modeling of {\it HST} imagery.

\end{abstract}


\section{Introduction}

Signatures of hot and cool starspots have been observed in numerous 
T~Tauri stars, primarily in the form of photometric variability, with 
{\it periodic} variability detected in hundreds of systems 
(e.g., Bouvier et al. 1993; Wichmann et al. 1998; Choi \& Herbst 1996; 
Makidon et al. 1997; Stassun et al. 1999).  
While cool spots have been mapped through Doppler imaging of weak-lined 
T~Tauri stars (e.g., Hatzes 1995), 
hot spots tend to be associated with variability in 
Classical T~Tauri stars (e.g., Kenyon et al. 
1994; Bouvier et al. 1993; Herbst et al. 1994). The hot spots are 
thought to be 
associated with the accretion process, and the currently favored magnetic 
accretion models naturally provide for hot spots on stellar surfaces 
(Ghosh \& Lamb 1979a, 1979b; K\"{o}nigl 1991; Shu et al. 1994; 
Ostriker \& Shu 1995; Najita 1995). 

Hot spots yield specific brightness and polarization variations as they 
rotate into and out of view 
(Wood et al. 1996; Mahdavi \& Kenyon 1998; Stassun \& Wood 1999).  
Wood \& Whitney (1998) investigated the effects of hot starspots on the 
morphology of scattered light disks.  The non-axisymmetric illumination of 
the disk by hot spots leads to an asymmetric brightening of the disk.  
Such asymmetric brightening has been detected in HST observations of the 
edge-on disk system HH30 (Burrows et al. 1996; Stapelfeldt et al. 1999a), 
hinting at possible agreement with the magnetic accretion model.

The gross consistency between the existing observations of HH30 
and the predictions of the magnetic accretion model is tantalizing,
but ambiguous due to the large timescales separating the HST observations. 
Periodic variability due to magnetic accretion is expected to 
match the rotational periods for Classical T~Tauri stars: 
of the order less than one day to tens of days 
(Attridge \& Herbst 1992; Bouvier et al. 1993; 
Edwards et al. 1993; Eaton et al. 1995; Stassun et al. 1999; Herbst, 
et al. 2000).  In some 
cases the hotspots are stable over many rotation periods.

Motivated by the apparent success of the magnetic accretion model at 
explaining the morphological variations observed in the limited HST imagery, 
we have undertaken a ground-based photometric monitoring campaign 
on HH30 to determine the timescale of any variability.  
In Section~2 we present the results of our observational campaign, in 
Section~3 we present models for the observed photometric variability, 
and we summarize our findings in Section~4.

\section{Observations}

Data for HH30 (RA = 04 31 37.5, DEC = 18 12 26.0;  $V\sim 19.5$, $I\sim 17.3$, 
Mundt \& Fried 1983) were obtained with the 1.2 m telescope at the
F. L. Whipple Observatory (FLWO), using the ``4Shooter'' CCD
mosaic (Szentgyorgyi et al. 2000) with four thinned, back-side
illuminated, AR coated Loral $2048^2$ CCDs. The camera has a pixel
scale of 0.335~arcsec/pixel and field of view of roughly 
11.5~arcmin for each chip.  The data were taken in the $2\times 2$
CCD binning mode.  
To obtain multi-wavelength variability information we used $VRI$
filters from the FLWO ``Harris Set''. We obtained data (see Figure~1), 
between September 7th 1999 and February 28th 2000,
usually consisting of up to three sets of 300~sec, 180~sec, 180~sec 
$VRI$ exposures. In total we have 24 useful images of
HH30 in the $V$-band and 27 useful images in each of the $RI$-bands,
spanning about 174~days. The seeing ranged from $FWHM=1.3$~arcsec
to about $4.0$~arcsec.  The $VRI$ data can be accessed from 
{\tt anonymous ftp} at {\tt ftp://cfa-ftp.harvard.edu/pub/kstanek/HH30}.

Preliminary processing of the CCD frames was done with the standard
routines in the IRAF-CCDPROC package.\footnote{IRAF is distributed by
the National Optical Astronomy Observatories, which are operated by
the Associations of Universities for Research in Astronomy, Inc.,
under cooperative agreement with the NSF} Photometric variability was
determined using two separate techniques: aperture photometry and
point-spread function (PSF) fitting.  PSF fitting was done using the
photometric pipeline of the project DIRECT (Kaluzny et al.~1998;
Stanek et al.~1998), based on the {\it Daophot/Allstar} photometry
package (Stetson 1987, 1991).  The differential light curves obtained
by both methods agreed well, which was reassuring considering
that the HH30 nebulosity is somewhat resolved in our images. As a
further check, we saw no correlation between the derived photometric
changes and the seeing or sky brightness.

The resulting $VRI$ differential light curves are shown in Figure~1 and
are adjusted so the faintest measurement in
each band corresponds to $\Delta$mag$=0$.  The HH30 nebulosity was observed
to vary in brightness over timescales of a few days by 
$\Delta V\sim 1.4\;$mag, $\Delta R\sim 1.0\;$mag, 
and $\Delta I \sim 1.1\;$mag. The changes in brightness are very well
correlated between the three photometric bands.  There are hints of 
periodicity with the rise and fall at the start of our observations and 
again around HJD 1570 in Figure~1.  
Independently for each band we searched for 
periodicity using a variant of the Lafler-Kinman (1965) string-length
technique, proposed by Stetson (1996) and also the Lomb (1976) normalized 
periodogram method.  The strongest signals in our periodogram techniques 
were found at 19.8~days (97\% probability) and 11.6~days (95\% probability).  
An 11.6~day period is more typical of T~Tauri stars and further monitoring 
of HH30 will possibly yield a more reliable period.  We emphasize that this
period detection is very tentative given the present data, and we
encourage follow-up observations to confirm or refute our finding.

Photometric variability in excess of 2~mags may be obtained by hot spots 
on the stellar surface (e.g., Mahdavi \& Kenyon 1998).  
Hot spots produce larger photometric 
variability at shorter wavelengths due to the increased spot/star 
luminosity ratio.  Our observations show that $\Delta V > \Delta I$, but 
$\Delta R < \Delta I$.  The apparent inconsistency of the $R$ band 
variability may be attributed to contamination from strong H$\alpha$ and 
[S~{\sc ii}] emission associated with HH30's jet (Burrows et al. 1996).  
Bouvier et al. (1999) discounted a hotspot model for AA~Tau because 
they did not detect any color variations ($\Delta$~mag$\sim1.4$ in all 
$BVRI$ bands).  Our HH30 data do show color variations and 
in the next section we investigate hot spot models for HH30 and 
restrict our simulations to the $V$ and $I$ bands due to the likely 
contamination of the $R$ band observations by jet emission.

\section{Models}

The large variability we observe may occur for a model with two 
diametrically opposed hot spots 
when the second spot is occulted (either by the star or the circumstellar 
disk) and the observed spot is seen directly at maximum light and is 
obscured by the star at minimum light.  However, for the high inclination 
of HH30 ($i>80^\circ$, Burrows et al. 1996) we only see a scattered 
light nebula and do not see the star directly.  For a model with two 
diametrically opposed hot spots, 
we found that at high inclinations the photometric variability was 
only around 0.2~mag (Wood \& Whitney 1998).  As 
one spot rotates behind the star, the second spot rotates to the front 
yielding the small amplitude variability.  In addition, because we are 
only detecting scattered light we always see some reflected light from the 
hot spots yielding small variations in the total intensity.  

The fact that we are viewing scattered light makes it 
impossible for us to apply inversion techniques to determine spot parameters 
(e.g., Bouvier et al. 1993; Vrba, Herbst, \& Booth 1988), 
since the light curves are sensitive not only to the star and spot 
parameters, but also the disk structure and circumstellar dust properties.  
The new WFPC2 observations presented by Stapelfeldt et al. (1999a) and the 
large variability present in Figure~1 suggests that one hot spot dominates 
the nebular and photometric variability.  
We now construct single hot spot models for HH30.

HH30 has been modeled by Burrows et al. 
(1996) and Wood et al. (1998) with a flared disk geometry,
\begin{equation}
\rho=\rho_0 \exp{ -{1\over 2} [z/h(\varpi )]^2  } / \varpi^\alpha 
\; ,
\end{equation}
where $\varpi$ is the radial coordinate in the disk midplane and the 
scale height increases with radius,
$h=h_0\left ( {\varpi /{R_\star}} \right )^\beta$.  
Following Burrows et al. (1996) and Wood et al. (1998) we adopt: 
$\beta = 9/8$, 
$\alpha=15/8$, $h_0=0.05R_\star$ (giving $h[100{\rm AU}] = 15$AU), and 
a disk mass of $2.5\times 10^{-4}M_\odot$.  
We construct scattered light models using our Monte Carlo 
radiation transfer code (Whitney \& Hartmann 1992; Wood \& Whitney 1998; 
Stassun \& Wood 1999) and assume there is one circular hot spot of 
radius $\theta_s$ at latitude $\phi_s$ on the stellar surface.  
The relative number of photons released from the spot and star is,
\begin{equation}
{ {N_s}\over{N_\star} } = { {1-\cos\theta_s}\over {1+\cos\theta_s} } 
{ {B_\lambda(T_s)}\over{B_\lambda(T_\star)} }  \; ,
\end{equation}
where we assume that the luminosities of the star and spot are Planck 
functions at the spot ($T_s$) and stellar ($T_\star$) temperatures.  

Figure~2 shows the simulated $V$ and $I$ band photometric variability for 
$\theta_s=20^\circ$, $\phi_s=65^\circ$, 
$T_s=10^4$K, $T_\star=3800$K (Wood \& Whitney 1998; Stapelfeldt et al. 1999).  
We assumed that the 
circumstellar dust can be characterized by a Kim, Martin, \& Hendry (1994, 
hereafter KMH) 
size distribution, typical of grains in the interstellar medium.  
For KMH grains the $V$ and $I$ band dust parameters (opacity, albedo, 
peak polarization, and asymmetry parameter in the Heyney-Greenstein 
scattering phase function) are displayed in Table~1.

Viewing this model at low inclinations we can easily obtain 
$\Delta V \sim 1.5$~mag.  However at $i=82^\circ$, we cannot reproduce the 
observed large amplitude variability --- we only see scattered light and 
always detect reflected light 
from the hot spot, irrespective of rotational phase.  Therefore 
the amplitude of variability is smaller ($\Delta V\sim0.5$~mag, 
$\Delta I\sim0.3$~mag) than when we see the star directly.  
One way of increasing the variability is 
to change the spot size and temperature.  Alternatively, if the forward 
beaming of the scattered light is increased (increasing the 
asymmetry parameter, $g$), we will see less scattered light from the spot 
at minimum light when it is occulted by the star.  
Such an increase in the asymmetry parameter, 
indicative of large grains in the disk, was found by Burrows et al. (1996).  
Other modeling efforts have also found evidence for larger, more forward 
throwing grains in protostellar environments (e.g., Lucas \& Roche 1998).  
We have therefore constructed models with larger, more forward throwing 
grains (see parameters in Table~1).  This population of 
grains was used by Cotera et al. (2000) and Whitney \& Wolff (2000) 
to model multiwavelength HST images of HH30.  
The dust grains are homogeneous spheres composed of either amorphous carbon
(BE1, Rouleau \& Martin 1991) or revised astronomical silicate (Weingartner 
\& Draine 2000).  The size distribution for each component is specified
using a power-law with exponential decay (i.e., $a^{-p} exp[-a/a_c]$)
with a cross-section weighted average radius (summed over both
components) of 0.092~\micron.  The relative numbers of each grain
type are such that the dust completely consumes slightly ``super-solar''
abundances in carbon and silicate: 400 ppm for C/H and 40 ppm for Si/H.  
We refer the reader to Whitney \& Wolff (2000) 
for a more detailed discussion of the dust model.

With the larger grains the 
amplitude of the variability is unaffected at low inclinations because 
the direct starlight is much brighter than the 
scattered light.  For $i=82^\circ$, we found larger variability 
($\Delta V\sim1$mag, $\Delta I\sim0.5$mag) than for the simulation using 
KMH grains due to the increased beaming of the scattered light by the forward 
throwing grains.  

Increasing the phase function asymmetry through larger grains increases 
the photometric variability, but it is still less than observed.  
Therefore, we have 
investigated other models which have larger spot/star luminosity ratios and 
different spot parameters.  
Figure~3 shows a simulation in which we obtain 
$\Delta V\sim 1.45$mags  and $\Delta I\sim 1.05$mags, in agreement with 
our observations.  This simulation has $T_s=10^4$K, $T_\star=3000$K, 
$\theta_s=20^\circ$, and $\phi_s=60^\circ$.  We can obtain the same 
amplitude of variability with a model that has $T_\star=3800$K and 
$T_s=2\times 10^4$K.  As the central star in the HH30 
system is not observed directly, its effective temperature 
is somewhat uncertain and may be cooler than the $T_\star=3800$K, M0 spectral 
type determined by Kenyon et al. (1998).  
The large spot/star luminosity, 
required to match our observations, could result in considerable veiling 
allowing for a later spectral type (see Kenyon et al. 1998).  
The spot size and latitude may be constrained 
through detailed modeling of time series images of the scattered light 
disk (Stapelfeldt, et al. 1999a).

\section{Summary}

We have presented $VRI$ photometric observations of HH30 that show it 
is highly variable over timescales of a few days ($\Delta V \sim 1.4$~mags, 
$\Delta R \sim 1.0$~mags, $\Delta I \sim 1.1$~mags), with hints of 
periodicity at 11.6 or 19.8~days.  This contrasts with the recent 
work of Stapelfeldt et al. (1999b) whose HST observations suggest a 
characteristic timescale on the order of 15~years, possibly related to 
inhomogeneities in the disk.  The much shorter timescale variability 
in our data indicates a stellar origin for the photometric variability 
we observe.  High resolution imaging is required to determine whether the 
variability we observe is related to the morphological variations in the 
scattered light images.

Comparing our findings to classical T~Tauri stars, HH30's variability is 
among the largest yet reported.  
We have modeled the variability 
in the context of the magnetospheric accretion model with a single hot spot 
on the stellar surface.  The large variability 
requires a large spot/star luminosity ratio and that the circumstellar 
grains are larger and more forward throwing than interstellar grains.  
Variability due to accretion hot spots is often stochastic 
(e.g., Herbst et al. 1994) 
and while our modeling has adopted a single stable hotspot, further 
observations are required to better sample the lightcurve and determine 
changes in the spot sizes and temperatures.  
Our modeling of the circumstellar dust is consistent with many other studies 
that are providing evidence for grain growth in dense environments of 
T~Tauri stars.

While our observations and modeling support the magnetic accretion hypothesis 
for HH30, alternative sources of the variability may be orbiting dust clouds, 
inhomogeneities in the disk structure close to the star, a binary star, and 
stellar flares.  Further monitoring of 
HH30 with sampling on the order of days will yield further insight into 
this system.  If the variability is due to magnetic accretion 
there will be specific time sequence variations of the phometry, 
polarimetry, and morphology of the scattered light disk due to 
illumination by hot spots.  Even if the accretion is stochastic there will 
be specific correlations between the photometry and polarimetry (Stassun \& 
Wood 1999, Fig.~7).  Rotational modulation will manifest itself in high 
resolution 
imaging studies through asymmetric brightening of scattered light images 
(Wood \& Whitney 1998; Stapelfeldt et al. 1999a).  Lower resolution imaging 
should still be able to detect the asymmetric brightening via shifting 
of the photometric centroid of the system as pointed out by Stapelfeldt et 
al. (1999a).

\acknowledgements

We thank Scott Kenyon for 
discussions relating to this project and the 
following people who helped us in obtaining observations of HH30 
at FLWO: Pauline Barmby, Ann Bragg, Paul Green, Saurabh Jha, Amy Mossman, 
Jose Munoz, and Rudy Schild.  The referee, Bill Herbst, provided 
comments and suggestions that have improved this paper.  
We acknowledge financial support from NASA's Long Term Space Astrophysics 
Research Program, NAG5~6039 (KW), NAG5~8412 (BW); the National Science 
Foundation, AST~9909966 (BW and KW); the HST Archival Research Program 
AR-08367.01-97A, (BW, KW, KS); and NASA contract NAS8-39073 (SJW).  
Support for KZS was provided by NASA through Hubble Fellowship grant
HF-01124.01-99A from the Space Telescope Science Institute, which is
operated by the Association of Universities for Research in Astronomy,
Inc., under NASA contract NAS5-26555.

\newpage

\begin{deluxetable}{lcccccc}
\tablenum{1}
\tablewidth{0pt}
\tablecaption{Parameters for Dust Grains}
\tablehead{
\colhead{Wave Band} & \colhead{$\kappa$ (cm$^2$/g)} &
\colhead{$a$} & \colhead{$g$} & \colhead{$P$}
}
\startdata
$V$(KMH)$\dotfill$  & 220 &  0.54 & 0.44 & 0.43 \\
$I$(KMH)$\dotfill$  & 105 &  0.49 & 0.29 & 0.70 \\
$V$(Large)$\dotfill$  & 174 &  0.47 & 0.61 & 0.38 \\
$I$(Large)$\dotfill$  & 134 &  0.49 & 0.58 & 0.34 \\
\enddata
\end{deluxetable}

\newpage

\begin{figure}[t]
\centerline{\plotfiddle{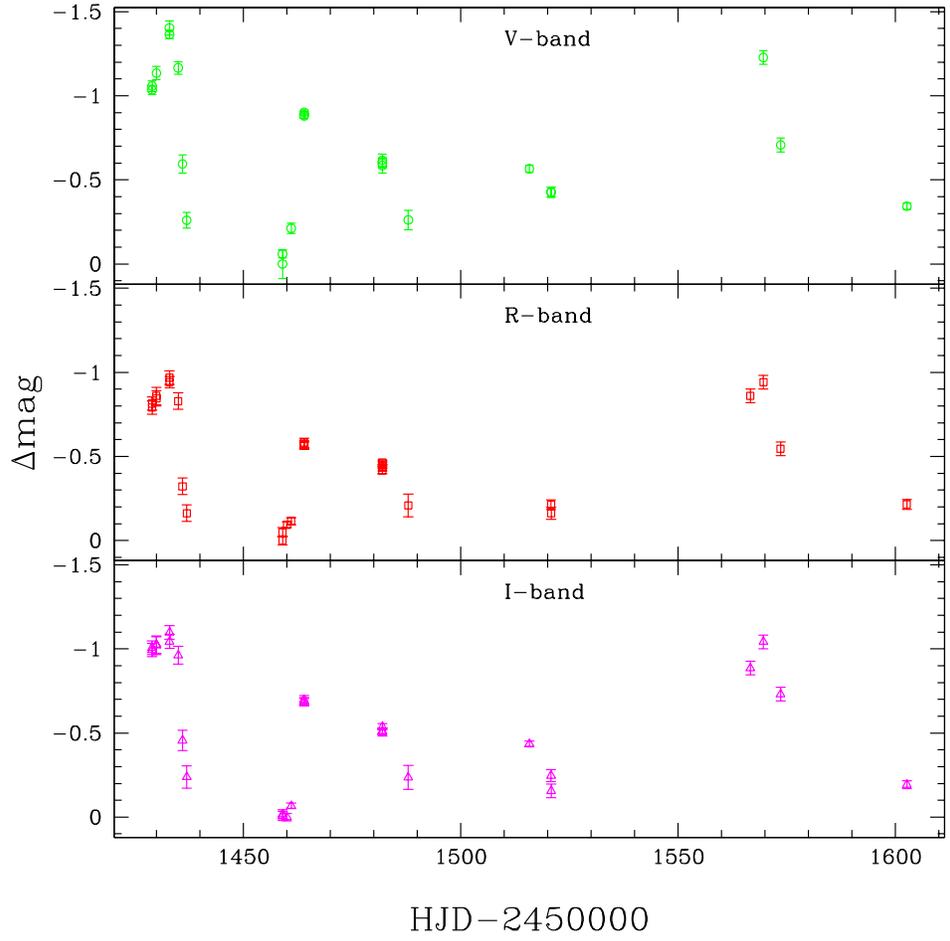}{6in}{0}{65}{65}{-420}{-20}}
\caption{HH30 $VRI$ photometric variability.}
\end{figure}

\begin{figure}[t]
\centerline{\plotfiddle{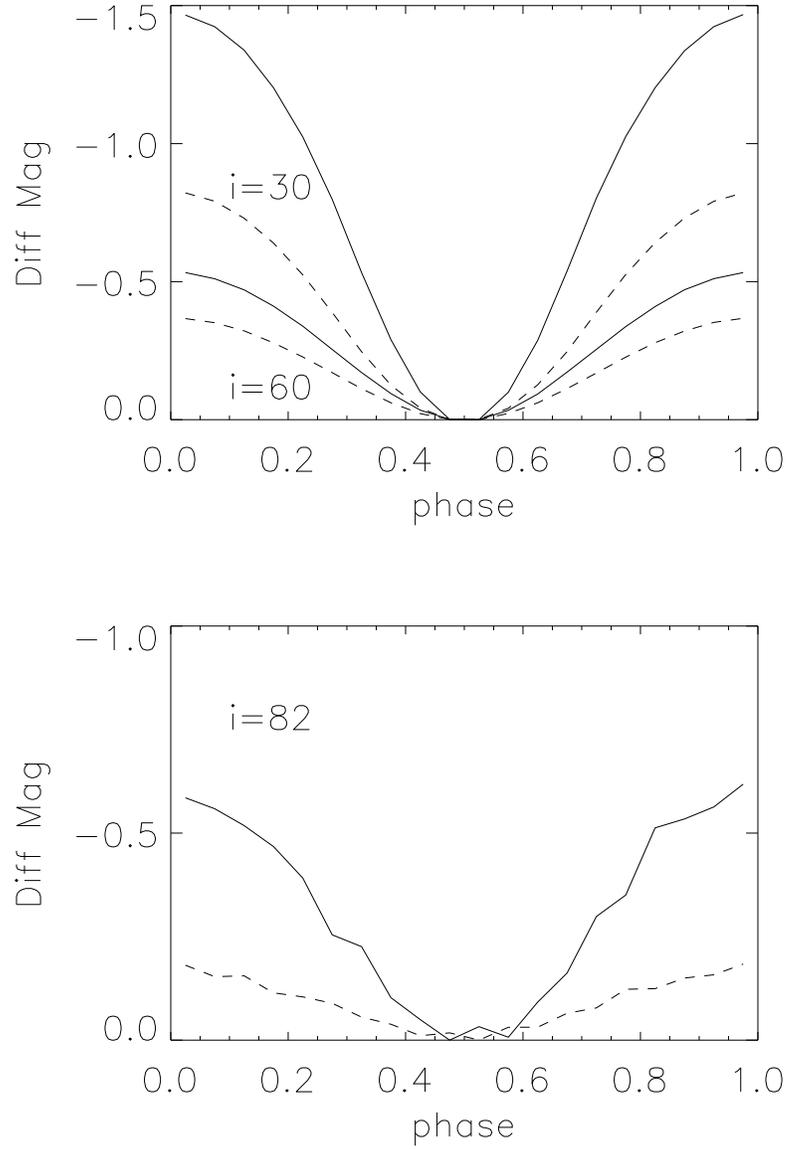}{6in}{0}{65}{65}{-420}{-20}}
\caption{Single spot model with KMH grains, $T_s=10^4$K, $T_\star = 3800$K, 
$\theta_s = 20^\circ$, $\phi_s = 65^\circ$. 
Upper panel shows the variability for viewing angles of $i=30^\circ$ and 
$i=60^\circ$, lower panel shows the variability for the more edge-on viewing 
angle $i=82^\circ$ appropriate to HH30.  
Solid line is the $V$ simulation and the dashed line is the $I$ simulation.}
\end{figure}

\begin{figure}[t]
\centerline{\plotfiddle{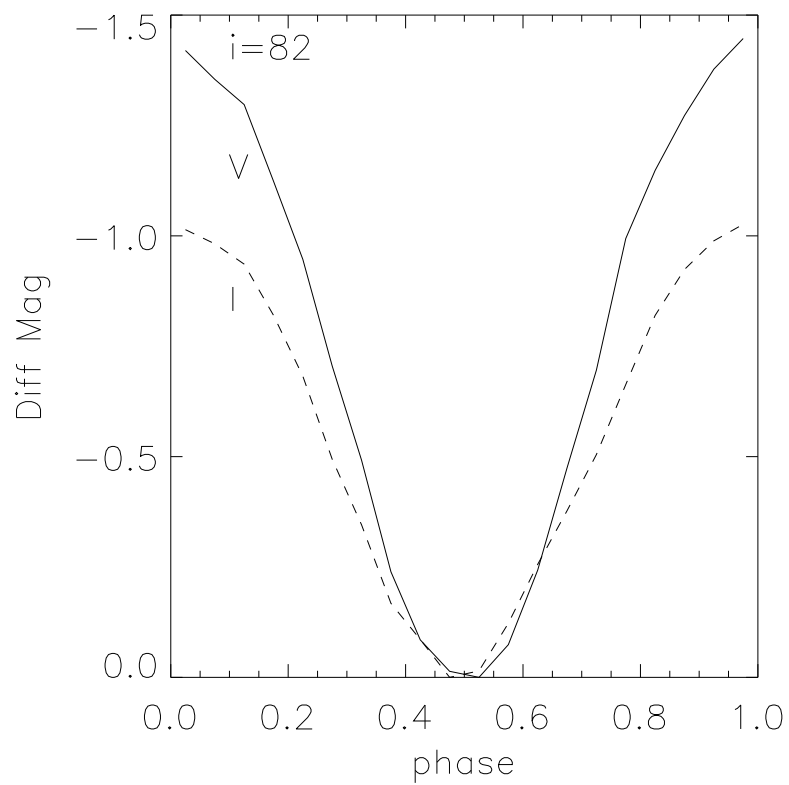}{6in}{0}{65}{65}{-420}{-20}}
\caption{Single spot model with larger, more forward throwing grains and 
a larger spot/star luminosity ratio, $T_s=10^4$K, $T_\star = 3000$K, 
$\theta_s = 20^\circ$, $\phi_s = 60^\circ$.  
Solid line is the $V$ simulation and the dashed line is the $I$ simulation.}
\end{figure}

\end{document}